

\magnification=1200
\vsize=7.3in
\hsize=5.4in
\def\no{\noindent}
\def\sy{\scriptscriptstyle}

\def\square
{\kern1pt\vbox{\hrule height
1.2pt\hbox{\vrule width 1.2pt\hskip 3pt
\vbox{\vskip 6pt}\hskip
3pt\vrule width 0.6pt}\hrule height 0.6pt}\kern1pt}

\def\sy{\scriptscriptstyle}
\def\dps{\displaystyle}
\def\D12{{\tau_1 - \tau_2}}
\def\D13{{\tau_1 - \tau_3}}
\def\D23{{\tau_2 - \tau_3}}
\def\half{{1\over 2}}
\def\pa{\partial}

\def\g12{{\dot {G_B}_{12}}}
\def\g13{{\dot {G_B}_{13}}}
\def\g23{{\dot {G_B}_{23}}}

\baselineskip1.0cm

\def\g{\mbox{g}}

\tolerance 10000
\baselineskip 13pt plus 1pt minus 1pt
\begingroup\nopagenumbers
\hskip 10cm HD-THEP-93-24\footnote{*}{Abridged version
(one footnote and one reference added)}
\vskip5pt
\centerline{{\bf ON THE CALCULATION OF
EFFECTIVE ACTIONS}}
\centerline{{\bf BY STRING METHODS}}
\vskip 36pt
\centerline{{Michael G. Schmidt and Christian Schubert
\footnote{**}{Partially supported by funds
provided by the Deutsche Forschungsgemeinschaft
}
\footnote{\dag}{E-mail address L56 @ DHDURZ1}}}
\vskip 12pt

\centerline{{Institut f\"ur Theoretische Physik}}
\centerline{Universit\"at Heidelberg}
\centerline{Philosophenweg 16}
\centerline{D-69120 Heidelberg}

\vskip 36pt
\centerline{\bf ABSTRACT}
{\medskip\narrower\noindent
Strassler's formulation of the string-derived Bern-Kosower formalism
is reconsidered with particular emphasis on effective actions and
form factors. Two- and three point form factors in the nonabelian
effective action are calculated and compared with those obtained in
the heat kernel approach of Barvinsky, Vilkovisky et al. We discuss the
Fock-Schwinger gauge and propose a manifestly covariant calculational
scheme for one-loop effective actions in gauge theory.

\medskip}

\eject
\bigskip
\endgroup
\pageno = 1

One of the main differences between string theory and particle theory
is the fact that in string theory
the full (on-shell) $S$-matrix can be calculated using the
Polyakov path integral, i.e. in first quantization.
As string theory reduces
to particle theory in the limit of infinite string tension, the same should
be possible in principle also for the particle S-matrix. And indeed, by a
painstaking
analysis of this limit, Bern and Kosower [1] have been able to
derive from string theory a new set of Feynman rules for the calculation
of one-loop amplitudes in conventional field theory which,
though quite different
from the conventional ones, appear to be completely equivalent [2].
Those rules
are particularly well-suited for calculations in gauge theories, as they
combine contributions of different Feynman diagrams into gauge
invariant structures [3].
Strassler [4] recently succeeded in deriving essentially the same set of
rules without explicit reference to string theory,
using one-dimensional path integrals with
Green functions adapted to the circle. Those path integrals are the
particle theory analogue of the genus one Polyakov path integral.
He further demonstrated
the advantages of the new formalism for the calculation of both abelian and
non-abelian effective actions [5].

In this paper, we will somewhat reformulate this method and apply it to
the computation of form factors, i.e. to the organization of the
higher derivative terms appearing in the effective action.
It will turn out that the two- and three- point
form factors for nonabelian gauge theory may be computed quite efficiently.
In the three-point case, the calculation
will be non-covariant. A manifestly covariant method of
calculation will be outlined and applied to the two-point case.

Let us begin with Strassler's [4] worldline path
integral expression for the
one-loop effective action of a Dirac spinor of mass $m$ minimally coupled
to an (abelian or non-abelian) background gauge field $A$:

$$\eqalign{
\Gamma\lbrack A\rbrack &= - 2 {\dps\int_0^{\infty}}
{dT\over T}
{\lbrack 2\pi\cal E T\rbrack}^{-{D\over 2}}
e^{-{{\cal E}\over 2}m^2T} {\cal N}
{\it Tr}{\dps\int} {\cal D} x\cal D\psi\cr
&{\phantom = }\times exp\Bigl [- \int_0^T d\tau
\Bigl (\Bigl ({1\over 2\cal E}\Bigr ) \dot x^2 + {1\over 2}\psi\dot\psi
+ igA_{\mu}\dot x^{\mu} - {1\over 2}ig{\cal E}
\psi^{\mu}F_{\mu\nu}\psi^{\nu}
\Bigr )\Bigr ].\cr
}\eqno(1)$$

\no
In this construction, the $x^{\mu}(\tau )$'s are the
periodic functions from the circle
with circumference $T$ into $D\;$- dimensional Euclidean spacetime,
and the $\psi^{\mu}(\tau )$'s their
antiperiodic supersymmetric partners, obeying

$$\lbrace\psi^{\mu},\psi^{\nu}\rbrace = g^{\mu\nu}. \eqno(2)$$

\no
$\cal N$ is the path integral normalization factor,

$${\cal N}^{-1} = {\it Tr}{\dps\int}
{\cal D}x{\cal D}\psi\; exp\Bigl [- \int_0^T d\tau
\Bigl (\Bigl ({1\over 2\cal E}\Bigr )\dot x^2 +
{1\over 2}\psi\dot\psi
\Bigr )\Bigr ],             \eqno(3)$$

\no
and $\cal E$ the worldline metric;
in the following we set ${\cal E} = 2$.
Note that our normalization includes the spinor degrees of freedom,
which makes for
a factor of $4$ compared to [4,5].
In the nonabelian case path ordering is implied.

\no
Insertions into the free path integral will be contracted using the
one-dimensional Green functions adapted to the (anti-)periodicity
conditions,

$$\eqalign{
\langle x^{\mu}(\tau_1)x^{\nu}(\tau_2)\rangle
   &= - g^{\mu\nu}G_B(\tau_1,\tau_2)
   = \quad - g^{\mu\nu}\biggl[ \mid \tau_1 - \tau_2\mid -
{{(\tau_1 - \tau_2)}^2\over T}\biggr],\cr
\langle \psi^{\mu}(\tau_1)\psi^{\nu}(\tau_2)\rangle
   &= {g^{\mu\nu}\over 2} G_F(\tau_1,\tau_2)
   = \quad {g^{\mu\nu}\over 2}{\rm sign}(\tau_1 -\tau_2 ).\cr
}\eqno(4)$$

\no

As a warm-up, let us reconsider the classical case of the 1-loop
effective action for spinor electrodynamics in a constant abelian
background [6]. Here choosing an appropriate gauge one may write
$A_{\mu} = {1\over 2}x^{\rho}F_{\rho\mu}$,
the full action remains quadratic,
and the calculation of $\Gamma [A]$ can be reduced to a calculation
of determinants as usual. However, in the present formalism those
determinants involve just integral
operators on the circle: Denoting
determinants by $\vert\vert$ and the derivative with respect to
$\tau$ by $\partial$, all we have to calculate is

$$\eqalign{
\Gamma\lbrack A\rbrack &= - 2
{\dps\int_0^{\infty}}{dT\over T}
{\lbrack 4\pi T\rbrack}^{-{D\over 2}} e^{-m^2T}{\cal N}
{\it Tr}{\dps\int} {\cal D} x\cal D\psi\cr
&{\phantom = }\times exp\Bigl [- \int_0^T d\tau
\Bigl (\Bigl ({1\over 4}\Bigr ) \dot x^2 + {1\over 2}\psi\dot\psi
+ {i\over 2}gx^{\mu}F_{\mu\nu}\dot x^{\nu}
- ig\psi^{\mu}F_{\mu\nu}\psi^{\nu}
\Bigr )\Bigr ]\cr
&= -2{\dps\int_0^{\infty}}{dT\over T}
{[4\pi T]}^{-{D\over 2}}e^{-m^2T}
{{\bigl\vert -{{\partial}^2\over 4}\bigr\vert}^{1\over 2}\over
{\bigl\vert -{{\partial}^2\over 4}+
{i\over 2}gF\otimes\partial\bigr\vert}
^{1\over 2}}
{{\bigl\vert{1\over 2}\partial\bigl\vert}^{-{1\over 2}}\over
{\bigl\vert{1\over 2}\partial - igF\bigr\vert}^{-{1\over 2}}}\cr
&= -2{\dps\int_0^{\infty}}{dT\over T}{[4\pi T]}^{-{D\over 2}}
e^{-m^2T}
{\bigl\vert{\bf I}- 2igF
\otimes \partial {{\partial}^2}^{-1}\bigr\vert}
^{-{1\over 2}}
{\bigl\vert{\bf I}-2igF
\otimes{\partial}^{-1}\bigr\vert}^{1\over 2}\cr
&= -2{\dps\int_0^{\infty}}{dT\over T}{[4\pi T]}^{-{D\over 2}}
e^{-m^2T}
{\bigl\vert{\bf I}-igF\otimes [{\dot G_B}]\bigr\vert}^{-{1\over 2}}
{\bigl\vert{\bf I}-igF\otimes [G_F]\bigr\vert}^{1\over 2}\cr
&= -2{\dps\int_0^{\infty}}{dT\over T}{[4\pi T]}^{-{D\over 2}}
e^{-m^2T}\cr
&\times
exp\Bigl [-{1\over 2}{\dps\sum_{n=1}^{\infty}}
{(-1)^{n+1}\over n}{(-ig)}^nT^n{\it Tr}F^n(Tr{[\dot G_B]}^n -
Tr{[G_F]}^n)\Bigr ].\cr
}\eqno(5)$$

\no
Here we used the $LogDet = TraceLog$\hskip3pt -- identity
in the last line, and with
$[G]$ we mean the integral operator
associated to the Green function $G$.
The $T$-- dependence of those integral
operators has been extracted by
a rescaling $\tau\to T\tau$, which leaves us with operators on the
unit circle.
Dots on a $G$ will always signify a derivative with respect to the
{\sl first} variable; in particular,

$$\eqalign{
\dot G_B(\tau_1 ,\tau_2 ) &= {\rm sign}(\tau_1 - \tau_2 )
- {2(\tau_1 - \tau_2)\over T},\cr
\ddot G_B(\tau_1,\tau_2 ) &=
2\delta (\tau_1 - \tau_2 ) - {2\over T}
.}\eqno(6)$$

\no
The operator traces may be computed using the Fourier bases
$\lbrace exp[2\pi ik\tau ]\vert k\in {\rm Z}\rbrace$
for the periodic and
$\lbrace exp[2\pi i(k+{1\over 2})\tau ]\vert k\in {\rm Z}\rbrace$
for the antiperiodic functions. The result is

$$\eqalign{
{\it Tr}{[\dot G_B]}^{2n} &= 2{(-1)}^n\zeta (2n){\pi}^{-2n},\cr
{\it Tr}{[G_F]}^{2n}      &= 2(2^{2n}-1){(-1)}^n\zeta (2n)
                            {\pi}^{-2n},
\cr
}\eqno(7)$$

\no
in terms of the Riemann $\zeta$-- function
(traces of odd powers vanish).
Finally

$$\Gamma [A] = -2{\dps\int_0^{\infty}}
{dT\over T}{[4\pi T]}^{-{D\over 2}}e^{-m^2T}
exp\Bigl [-{\dps\sum_{n=1}^{\infty}}
{{(gT)}^{2n}\over n}(2^{2n-1}-1)\zeta (2n)
{\pi}^{-2n}{\it Tr}F^{2n}\Bigr ].
\eqno(8)$$

\no
This is the generating functional for eq. (3.23) in [5]
\footnote{*}{We believe that
the recursion relations derived in [5] for the above traces should
be related to the well-known recursion
formula for the Bernoulli numbers.}
\footnote{**}{Analogous methods have been already used
by Metsaev and
Tseytlin [15] to obtain a generalized Schwinger formula for the
bosonic string and, by taking the infinite string tension limit
of this generalization, for YM theory.}.
It may be transformed into the standard representation
(see e.g. [7])

$$\Gamma[A] = {1\over 8\pi^2}{\dps\int_0^{\infty}}
{ds\over s}e^{-ism^2}
\biggl[g^2ab{cosh(gas)cos(gbs)\over
sinh(gas)sin(gbs)}-{1\over s^2}\biggr ],
\eqno(9)$$

\no
where

$$\eqalign{
a^2 &= \half\biggl [{\bf E}^2 - {\bf B}^2 + \sqrt{{({\bf E}^2 -
{\bf B}^2)}^2+4{({\bf E}\cdot{\bf B})}^2}\biggr]\cr
b^2 &=  \half\biggl [-({\bf E}^2 - {\bf B}^2) + \sqrt{{({\bf E}^2 -
{\bf B}^2)}^2+4{({\bf E}\cdot{\bf B})}^2}\biggr],\cr
}\eqno(10)$$

\no
by diagonalizing the matrix $F_{\mu\nu}$, showing

$${\it Tr}F^{2n} = 2\Bigl[{(a^2)}^n +
{(-b^2)}^n\Bigr], \eqno(11)$$

\no
and using the Taylor expansions of $ln[xcot(x)],ln[xcoth(x)].$

After this amusing calculation, let us proceed to our main objective,
which is the calculation of form factors in the nonabelian theory.
That means we want to calculate
(taking the case of a (complex) scalar loop first)

$$
\Gamma\lbrack A\rbrack = {\dps\int_0^{\infty}}
{dT\over T}
{\lbrack 4\pi T\rbrack}^{-{D\over 2}}e^{-m^2T} {\cal N}
{\it Tr}{\dps\int} {\cal D} x\;
exp\Bigl [- \int_0^T d\tau
\bigl (  {1\over 4}      {\dot x}^2
+ igA_{\mu}\dot x^{\mu}
\bigr )\Bigr ]
\eqno(12)$$

\no
up to a fixed power in the background field, but to all orders
in the covariant derivatives $D_{\mu} = \partial_{\mu} -igA_{\mu}$
of the field (for the history of this subject and for
possible physical applications see ref. [8] and references therein).
To begin with, we introduce the loop center of mass $x_0$,
writing

$$x^{\mu}(\tau) = x^{\mu}_0  +  y^{\mu} (\tau )\eqno(13)$$

\no with

$$\int_0^T d\tau\,   y^{\mu} (\tau ) = 0,\eqno(14)$$

\no
and extract the integral over the center of mass from
the path integral:

$${\dps\int}{\cal D}x =
{\dps\int}d^D x_0{\dps\int}{\cal D} y.\eqno(15)$$

\no
Next we Taylor expand the interaction part
$\dot x^{\mu}A_{\mu}(x(\tau ))$
with respect to $x_0$, use ${\dot x}^{\mu} = {\dot y}^{\mu}$ to write

$$\dot x^{\mu}A_{\mu}(x) =
{\dot y^{\mu}}e^{y \partial}A_{\mu}(x_0)\eqno(16)$$

\no
and expand the path-ordered interaction exponential to get

$$\eqalign{
\Gamma\lbrack A\rbrack &= {\it Tr}{\dps\int_0^{\infty}}
{dT\over T}{\lbrack 4\pi T\rbrack}^{-{D\over 2}}e^{-m^2T}
{\dps\int}d^D x_0
{\dps\sum_{n=0}^{\infty}}{{(-ig)}^n\over n}
T\int_0^{\tau_1 = T}d\tau_2\int_0^{\tau_{2}} d\tau_3
\ldots\int_0^{\tau_{n-1}} d\tau_n\cr
&\times{\cal N}
{\dps\int} {\cal D} y
{\dot y^{\mu_1}}(\tau_1 )e^{y({\scriptscriptstyle\tau_1})
\partial_{(1)}}A^{(1)}_{\mu_1}(x_0)
\ldots
{\dot y^{\mu_n}}(\tau_n )e^{y({\scriptscriptstyle\tau_n})
\partial_{(n)}}A^{(n)}_{\mu_n}(x_0)
exp \Bigl [- \int_0^T d\tau
  {{\dot y}^2\over 4}\Bigr ].
}\eqno(17)$$

\no
We have labeled the background fields $A_{\mu_1},\ldots,A_{\mu_n}$,
and the first $\tau\,$ -- integration has
been eliminated by using the
freedom of choosing the point $0$ somewhere on the loop.
Instead of restricting the colour trace by fixing the cyclic order,
we have introduced an explicit factor of ${1\over n}.$
The single terms in this expansion may
now be computed by Wick contractions
in the one-dimensional worldline field theory, using the above bosonic
Green function on the circle and formulas familiar from string theory,

$$\eqalign{
\langle e^{y({\sy\tau_1})\partial_{(1)}}
e^{y({\sy\tau_2})\partial_{(2)}}\rangle &=
e^{{\scriptstyle -G_B}({\sy\tau_1},
{\sy\tau_2})\partial_{(1)}\partial_{(2)}},\cr
\langle \dot y^{\mu}({\tau_1})
e^{y({\sy\tau_2})\partial_{(2)}}\rangle &=
-\dot G_B({\tau_1},{\tau_2})\partial_{(2)}^{\mu}.\cr
}\eqno(18)$$

\no
For the term quadratic in $A_{\mu}$, this leads to a single
$\tau\;$ -- integration

$$T\int_0^Td\tau_2\Bigl [\ddot G_B(T,{\tau_2})g^{\mu_1\mu_2}
   -{\dot G_B(T,{\tau_2})}^2\partial_{(2)}^{\mu_1}
   \partial_{(1)}^{\mu_2}\Bigr ]e^{-G_B({\sy T},
{\sy\tau_2})\partial_{(1)}\partial_{(2)}}
A_{\mu_1}^{(1)}A_{\mu_2}^{(2)}
,\eqno(19)$$

\no
which using the scaling properties of $G_B,\dot G_B,\ddot G_B$
and partial integration with respect to $\tau_2$ may be transformed
into

$$T^2\int_0^1 du_2{\dot G_B(1,{u_2})}^2
\bigl [g^{\mu_1\mu_2}\pa_{(1)}{\pa_{(2)}}-\partial_{(2)}^{\mu_1}
   \partial_{(1)}^{\mu_2}\bigr ]e^{-TG_B({\sy 1},
{\sy u_2})\partial_{(1)}\partial_{(2)}}
A_{\mu_1}^{(1)}A_{\mu_2}^{(2)}
,\eqno(20)$$

\no
where $G_B$ now refers to the unit circle,
$G_B(1,u_2 ) = u_2(1-u_2).$
But

$$[g^{\mu_1\mu_2}\pa_{(1)}\pa_{(2)}
-\partial_{(2)}^{\mu_1}\partial_{(1)}^{\mu_2}\bigr ]A^{(1)}_{\mu_1}
A^{(2)}_{\mu_2}$$

\no
is just the "abelian" part of
$\half F^{(1)}_{\mu\nu}F^{(2) \mu\nu}$,
and we need not see more of this tensor to determine its form factor:
Anticipating covariantization and performing
further partial integrations both with respect to $u_2$
(to get rid of the $\dot G_B^2$)
and to $x_0$ (yielding
$\partial_{(1)}\partial_{(2)} = - {\partial_{(2)}}^2)$
to get an expression standardized in the sense of [8], we can already
predict the following result for the
full two-point one-loop effective action:

$$\eqalign{
\Gamma_2\lbrack A\rbrack &= {\it Tr}{\dps\int_0^{\infty}}
{dT\over T}{\lbrack 4\pi T\rbrack}^{-{D\over 2}}e^{-m^2T}
T^2 (-g^2)
{\dps\int}d^D x_0
F_{\mu\nu}(x_0)F_2^{scal}(\xi )F^{\mu\nu}(x_0)
}\eqno(21)$$

\no
with
$$\eqalign{
F_2^{scal}(\xi ) &= \biggl [-\half {f(\xi ) - 1\over\xi}\biggr ],\cr
f(\xi ) &= {\dps\int_0^1}du e^{-u(1-u)\xi},\cr
 \xi &=  -T D_{\mu}D^{\mu}.\cr
}\eqno(22)$$

\no
This is the same expression as has been reached by [8] using
covariant perturbation theory.

Before proceeding to the three-point case,
let us say something about the
systematics of the calculation in general.
Every $\dot y(\tau_i)$ has to be
contracted, either with another $\dot y(\tau_k)$,
or with an $e^{y({\sy\tau_i})\partial_{(i)}}$.
The most basic case is if there
are only contractions of the second kind,
as the result will be just one
or a product of several closed chains of    $G_B$' s,

$$\dot G_B(\tau_{i_1},\tau_{i_2})\dot G_B(\tau_{i_2},
\tau_{i_3})\ldots
\dot G_B(\tau_{i_n},\tau_{i_1})
\pa_{(i_1)}^{\mu_{i_2}}\ldots\pa_{(i_n)}^{\mu_{i_1}},\eqno(23)$$

\no
multiplied by a universal factor of

$$exp\Bigl [-{\sum_{i<k}G_B({\tau_i},
{\tau_k})\pa_{(i)}\pa_{(k)}}\Bigr ]\eqno(24)$$

\no
from the contraction of the exponentials among themselves.

\no
Contractions of the first kind would result in $\ddot G_B$'s, which
might be reverted into $\dot G_B$'s by partial integrations later
(this problem has already been discussed at length in refs. [1-5]).
{}From the systematic point of view, however, we prefer to
eliminate the $\ddot G_B$'s {\sl before} they arise:
Before contracting
a $\dot y(\tau_i)$ with a $\dot y(\tau_k)$, one performs a partial
integration in $\tau_k$. This leaves a new $\dot y(\tau_k)$ after
contraction of the $\dot y(\tau_i)$ in the main part of the partial
integration, and hence
the $\langle\dot y y\rangle$ chain construction works
in this case, too. But there are further terms both from the
boundary of the $\tau_k$ integration and from differentiating
the $\tau_k$-- dependent boundary of $\tau_{k+1}$.

\no
Altogether we obtain for one contraction step of the first
kind a factor of\footnote{*}{($\delta\;$-- functions with
undefined $\tau_{k\pm 1}$ should
be skipped)}

$$\eqalign{
&-\dot G_B(\tau_i,\tau_k)
g^{\mu_i\mu_k}\bigl[ - \dot y^{\lambda}(\tau_k)
\pa^{(k)}_{\lambda} + \delta (\tau_k - \tau_{k-1}) - \delta (\tau_k)
-\delta (\tau_k - \tau_{k+1})\bigr ].\cr
}\eqno(25)$$

\no
Whereas the first term of this expression
is part of a continuing chain, the other ones
may be represented pictorially as the end of a
$\langle\dot y y\rangle$ chain
at $\tau_k$ ($\ne \tau_{i_0}$, if
contraction has started with  $\dot y(\tau_{i_0})$ ).
As one can infer already from the two-point calculation, it
is useful to always combine the
first term with the double contraction

$$\langle\dot y(\tau_i)e^{y({\sy\tau_k})\partial_{(k)}}\rangle
\langle e^{y({\sy\tau_i})\partial_{(i)}}\dot y(\tau_k)\rangle.$$

\no
The result of one particular total contraction will be a product of
both open and closed chains of $\dot G_B$'s, where open chains
start freely and may but need not necessarily
end on other chains (closed or open).

\no
Treating the three-point path integral in this way, we obtain
(after the usual rescaling $\tau = Tu$) the $u\;$-- integrals

$$\eqalign{
&{\dps\int_0^1 du_2 \int_0^{u_2}du_3}\Bigl \lbrace
\Bigl [ (g^{\lambda\mu}\pa_{\sy (1)}\pa_{\sy (2)} -
\pa_{\sy (2)}^{\lambda}\pa_{\sy (1)}^{\mu} )
\dot {G_B}_{12}^2 (\pa_{\sy (1)}^{\nu}
\dot {G_B}_{13}+\pa_{\sy (2)}^{\nu}
\dot {G_B}_{23}) + {\rm \; cyclic\; terms\;}\Bigr ]\cr
& - \dot {G_B}_{12}\dot {G_B}_{13}\dot {G_B}_{23}\Bigl [
g^{\lambda\mu}(\pa_{\sy (1)}^{\nu}\pa_{\sy (2)}\pa_{\sy (3)}
 - \pa_{\sy (2)}^{\nu}\pa_{\sy (1)}\pa_{\sy (3)})
+  {\rm\; c.\; t.\;}
+ \pa_{\sy (3)}^{\lambda}\pa_{\sy (1)}^{\mu}\pa_{\sy (2)}^{\nu}
- \pa_{\sy (2)}^{\lambda}
\pa_{\sy (3)}^{\mu}\pa_{\sy (1)}^{\nu}\biggr ]\Bigr\rbrace\cr
&\times exp{\Bigl[-{G_B}_{12}\pa_{\sy (1)}
\pa_{\sy (2)}-{G_B}_{13}\pa_{\sy (1)}\pa_{\sy (3)}
 -{G_B}_{23}\pa_{\sy (2)}\pa_{\sy (3)}\Bigl]}
A^{(1)}_{\lambda}(x_0)A^{(2)}_{\mu}(x_0)A^{(3)}_{\nu}(x_0)\cr
& +{\rm \; boundary\; terms\quad},\cr
}\eqno(26)$$

\no
where "boundary terms" stands for all terms involving one of the
$\delta\;$-- functions in the contraction formula (25).
Those terms reduce
to a single $u\;$-- integration and may be written as functions of
$f(\pa_{(1)}^2),f(\pa_{(2)}^2)$ and $f(\pa_{(3)}^2)$,
where $f$ is the basic two-point
function defined in eqs.(22)
(part of them in fact does not contribute
to the three-point form factors, but to the covariantization of the
two-point form factor).
According to the analysis of [8], on the three-point level there
are -- up to powers of covariant derivatives, Bianchi
identities and partial integrations in $x_0\;$-- two independent
invariant tensor
structures, which may be chosen as

$$\eqalign{
 T_1 &=   {\rm Tr}\,
F_{\mu}^{\lambda}F_{\lambda}^{\nu}F_{\nu}^{\mu},\cr
T_2 &=
{\rm Tr}\,F^{\alpha\beta}D^{\mu}F_{\mu\alpha}
D^{\nu}F_{\nu\beta}.\cr
}\eqno(27)$$

\no
The terms in the second square bracket of
formula (26) combine to produce
simply the "abelian" part of $-T_1$,
while the terms in the first square bracket contribute to both
$T_1$ and $T_2$. Collecting all contributions to $T_1$ we obtain
for its coefficient the $u\;$-- integral

$$\eqalign{
{\dps\int_0^1 du_2}&{\int_0^{u_2}du_3}\,
 exp{\Bigl[-{G_B}_{12}\pa_{\sy (1)}
\pa_{\sy (2)}-{G_B}_{13}\pa_{\sy (1)}\pa_{\sy (3)}
 -{G_B}_{23}\pa_{\sy (2)}\pa_{\sy (3)}\Bigr]}
\biggl\lbrace
  \dot {G_B}_{12}\dot {G_B}_{13}\dot {G_B}_{23}\cr
& -\half\Bigl[ \dot {G_B}_{12}^2 (-\dot{G_B}_{13}+\dot{G_B}_{23})
+ \dot {G_B}_{13}^2 (\dot{G_B}_{12}+\dot{G_B}_{23})
+ \dot {G_B}_{23}^2 (\dot{G_B}_{12}-\dot{G_B}_{13})\Bigr]
\biggl\rbrace .\cr
}\eqno(28)$$

\no
To compare this coefficient with the results of [8], we again
standardize by partially integrating both with
respect to $u$ and $x_0$, introduce

$$\xi_i = -TD^{(i)\mu}D^{(i)}_{\mu},$$

\no
and performe the change of variables

$$\eqalign{
\alpha_1 &= u_2-u_3\cr
\alpha_2 &= 1-(u_1-u_3)\cr
\alpha_3 &= u_1-u_2,\cr
}\eqno(29)$$

\no
leading to

$$\eqalign{
\Gamma_{\sy{T_1}}\lbrack A\rbrack = {\dps\int_0^{\infty}}
{dT\over T}{\lbrack 4\pi T\rbrack}^{-{D\over 2}}e^{-m^2T}
T^3(-ig^3){\dps\int}d^D x_0
F^{scal}_{\sy{T_1}}(\xi_1,\xi_2,\xi_3)
{\rm Tr}\,F_{\mu}^{\lambda}F_{\lambda}^{\nu}F_{\nu}^{\mu}(x_0),\cr
F_{T_1}^{scal}(\xi_1,\xi_2,\xi_3)
={4\over {3\Delta^2}}
\Bigl[ \Sigma (\Delta + \Xi ) + 12\Xi + 8{\Xi^2\over\Delta}\Bigr]
F(\xi_1,\xi_2,\xi_3) + {\rm boundary\; terms\quad}.
}\eqno(30)$$

\no
Here we have defined

$$\eqalign{
\Sigma &= \xi_1 + \xi_2 + \xi_3\cr
\Xi    &= \xi_1\xi_2\xi_3\cr
\Delta &= \xi_1^2+\xi_2^2+\xi_3^2
          -2\xi_1\xi_2-2\xi_1\xi_3-2\xi_2\xi_3\cr
}\eqno(31)$$

\no
and the basic three-point function

$$\eqalign{
F(\xi_1,\xi_2,\xi_3) =
{\dps\int_{\alpha\ge 0}}d^3\alpha\,
\delta (1-\alpha_1-\alpha_2-\alpha_3)
 exp[-\alpha_1\alpha_2\xi_3-\alpha_2\alpha_3\xi_1-
\alpha_3\alpha_1\xi_2 ]\cr
}\eqno(32)$$

\no
originating from the universal contraction of exponentials alone.
Again
we have assumed that the higher order calculations will produce
the missing "nonabelian" parts of the tensor $T_1$,
as should be granted by the
consistency of the Bern-Kosower formalism.
Expression (30) for the
form factor $F_{\sy{T_1}}^{scal}$ of $T_1$
now may be easily identified with the
one found in [8] by covariant perturbation theory
\footnote{*}{The
fact that superparticle path integrals
can be an efficient alternative
to heat kernel methods is well-known from the calculation of
index densities [9].}.

Now let us return to the case of the spinor loop, eq. (1).
At the two-point level,
there is one additional contraction to compute compared to the
scalar case, namely

$$\langle\psi^{\mu_1}(\tau_1)e^{y({\sy\tau_1})\partial_{(1)}}
F^{(1)}_{\mu_1\nu_1}(x_0)\psi^{\nu_1}(\tau_1)
\psi^{\mu_2}(\tau_2)e^{y({\sy\tau_2})\partial_{(2)}}
F^{(2)}_{\mu_2\nu_2}(x_0)\psi^{\nu_2}(\tau_2)\rangle,\eqno(33)$$

\no
giving a $u\;$-- integral

$${1\over 4}T^2\int_0^1 du_2{G_F(1,{u_2})}^2
\bigl [g^{\mu_1\nu_2}g^{\nu_1\mu_2}-g^{\mu_1\mu_2}g^{\nu_1\nu_2}\bigr ]
   e^{-TG_B({\sy u_1},{\sy u_2})\partial_{(1)}\partial_{(2)}}
F^{(1)}_{\mu_1\nu_1}F^{(2)}_{\mu_2\nu_2}.
\eqno(34)$$

\no
Using $G_F(1,{u_2}) = 1$ and taking the global factor of $-2$ in (1) into
account, we obtain the following relationship between the
scalar and spinor two-point form factors:

$$F_2^{spin}(\xi ) = -2 [F_2^{scal}(\xi )-{1\over 4} f(\xi )].\eqno(35)$$

\no
In the three-point case, we have two additional contraction possibilities:
The first one is a contraction of

$$\langle\psi^{\mu_1}e^{y({\sy\tau_1})\partial_{(1)}}
F^{(1)}_{\mu_1\nu_1}\psi^{\nu_1}
\psi^{\mu_2}e^{y({\sy\tau_2})\partial_{(2)}}
F^{(2)}_{\mu_2\nu_2}\psi^{\nu_2}
\psi^{\mu_3}e^{y({\sy\tau_3})\partial_{(3)}}
F^{(3)}_{\mu_3\nu_3}\psi^{\nu_3}\rangle,\eqno(36)$$

\no
and it simply produces a contribution of
${1\over 3}F(\xi_1,\xi_2,\xi_3)$ to $T_1$.
The second one,

$$\langle\psi^{\mu_1}e^{y({\sy\tau_1})\partial_{(1)}}
F^{(1)}_{\mu_1\nu_1}\psi^{\nu_1}
\psi^{\mu_2}e^{y({\sy\tau_2})\partial_{(2)}}
F^{(2)}_{\mu_2\nu_2}\psi^{\nu_2}
\dot y^{\mu_3}e^{y({\sy\tau_3})\partial_{(3)}}A_{\mu_3}
\rangle,\eqno(37)$$

\no
contributes to $T_1$ and $T_2$, with a contribution to $T_1$ of
$-{1\over 3}F(\xi_1,\xi_2,\xi_3)$ and thus cancelling the first one.
\no
The final result for the $T_1\;$-- form factor for spinor loops
becomes simply

$$F_{\sy{T_1}}^{spin}(\xi_1,\xi_2,\xi_3) =
-2 F_{\sy{T_1}}^{scal}(\xi_1,\xi_2,\xi_3).
\eqno(38)$$

\no
The form factors for $T_2$ will be treated in a more detailed
publication [10],
as well as those of the cyclic four-point tensor.
Now let us rather sketch briefly how this type of
calculations might be done in a manifestly covariant way.

\no
For fixed $x_0$ the Fock--Schwinger gauge [11]

$$ y^\mu A^a_\mu(x_0+y(\tau))=0\eqno(39)$$

\no
is very convenient for writing down manifestly gauge-invariant
expressions for the effective action. In this gauge,

$$A^a_\mu(x_0+y)=\int_0^1 d\eta\eta F^a_{\rho\mu}
(x_0+\eta y)y^\rho.\eqno(40)$$

\no
Following the proof of [11], $F_{\rho\mu}^a$ can be
{\sl covariantly} Taylor--expanded as

$$F^a_{\rho\mu}(x_0+\eta y)=(e^{\eta y\cdot D})^{ab}F^b
_{\rho\mu}(x_0)\eqno(41)$$

\no
leading to

$$\eqalign{
A_\mu(x_0+y)&=\int^1_0 d\eta\eta\, y^\rho e^{\eta y\cdot D}
F_{\rho\mu}(x_0)\cr
&=\half y^\rho F_{\rho\mu}+{1\over 3}y_{\nu}y_{\rho} D_\nu
F_{\rho\mu}+...\cr
}\eqno(42)$$

\no
Expanding again our starting expression (12)
for the scalar effective action
in powers of $g$ and using the Fock-Schwinger gauge relation (42),
we arrive at

$$\eqalign{
\Gamma [F] &=
{\dps\int_0^{\infty}}
{dT\over T}{\lbrack 4\pi T\rbrack}^{-{D\over 2}}e^{-m^2T}
{\dps\int}d^D x_0
{\dps\sum_{n=0}^{\infty}}{{(-ig)}^n\over n}
T\int_0^{\tau_1 = T}d\tau_2\int_0^{\tau_{2}} d\tau_3
\ldots\int_0^{\tau_{n-1}} d\tau_n\cr
&\times\int^1_0 d\eta_1\eta_1...\int^1_0 d\eta_n\eta_n{1\over {\eta_1
...\eta_n}}{{\partial}\over{\partial D^{(1)}_{\rho_1}}}...{{\partial}
\over {\partial D^{(n)}_{\rho_n}}}\cr
&{\cal N}{\dps\int {\cal D}y}\dot y^{\mu_1}(\tau_1)
e^{\eta_1y(\tau_1)\cdot D^{(1)}}\ldots\dot y
^{\mu_n}(\tau_n)e^{\eta_ny(\tau_n)\cdot D^{(n)}}
exp \Bigl [- \int_0^T d\tau
  {{\dot y}^2\over 4}\Bigr ]\cr
&F^{a_1}_{\rho_1\mu_1}...F^{a_n}_{\rho_n\mu_n}Tr(T^{a_1}...T^{a_n}),
\cr
}\eqno(43)$$

\no
where the $D^{(i)}$ act on $F^{a_i}_{\rho_i\mu_i}$ and in this sense
commute in the prefactor.
${1\over \eta_i}{{\partial}\over{\partial D^{(i)}_{\rho_i}}}$ creates
the $y^{\rho_i}$ of
relation (42).
Note that the contraction
of a certain $y(\tau_i)$ in $e^{\eta_i y(\tau_i)D^{(i)}}$ with
$y(\tau_k)(k\not=i)$ and the commutation of
$y(\tau_i) D^{(i)}$ factors in the
exponential are noncommuting operations in the nonabelian case.

The systematics of the contractions are the same as before. However,
the contraction of the exponentials now leads to

$$ exp\Bigl [-{\sum_{i<k}\eta_i\eta_k
G_B(\tau_i,\tau_k)D^{(i)}D^{(k)}}\Bigr ],\eqno(44)$$

\no
which is symbolical in the case of nonabelian gauge theories:
Each polynomial in $D^{(i)}_\mu$ for
fixed $i$ has to be written in all
possible orderings, and the resulting expression has to be normalized
by the number of possible orderings.
This is also the way further factors $D^{(i)}$ have to be handled.

In nonabelian gauge theory the ordering of $T^a$ matrices and of the
various covariant $D^{(i)}$ acting on the $F_{\mu\nu}$
in the ith position requires reduction to a set of
standard invariants using $D$-commutators
(leading to further $F_{\mu\nu}$'s) and
Bianchi identities, an admittedly painful procedure.

Formfactors of $n$ fields $F_{\mu\nu}$
in the covariant formalism just
described contain one further integration
$\eta_i$ per field. In the case
of two fields $F_{\mu\nu}$ we have checked
that the additional integrations
can be performed and that one arrives
at the $\alpha\;$-- form of ref. [8].
In the case with three $F_{\mu\nu}$ this seems to be a
more serious problem and is under consideration.

For application in QCD and in the standard
electroweak theory of course besides
the complex scalar fields we need gauge
fields and fermionic fields in the loop
and also Higgs fields in the background.
In this context the above representation of the fermions by
spinning superparticles in the loop [12]
appears to be very economical.
However if we want to calculate fluctuation determinants,
e.g. for a discussion of radiative corrections
to instantons, sphalerons, bounce identities,
a different technique seems to have advantages:
To calculate the bilinear part
of the quantum Lagrangian in a gauge field -- Higgs fields background
in a convenient quantum gauge
(for the background field we can have a separate gauge).
The 't Hooft-Feynman background gauge seems to be most
practical - see ref. [13,14].
The corresponding set of operators has the form $-D^2(A)+V
(\phi,F_{\mu\nu}),$ with $D(A)$
being some collection of covariant derivatives
in some gauge group representations and $V$
a general potential matrix built
out of background scalar field $\phi$, field strength $F_{\mu\nu}$,
and covariant derivatives thereof
(in the case of the electroweak theory the main matrix
operator depends on $3\times 3+4=13$ fields [13]). $LogDet(-D^2+V)$
is then evaluated  with the methods discussed above.
Proceeding as in the pure gauge case, there will
be also factors
$e^{y(u_i)\cdot D^{(i)}}V^{(i)}(\phi,F_{\mu\nu})$ in the trace.
The resulting expressions are lengthy but straightforward to write
down and to evaluate for a finite number of $V$ and $D$,
if a convention for a set of
independent operators is fixed.
This procedure does not have the cancellations
of that of ref. [13] with plane wave insertions in $TrLog$,
and it is within
the range of the method to extend their results beyond the third order
\footnote{*}{(work in progress)} (or beyond the 6th order in the
case without background field [14]).

\vskip.5cm
\no
We would like to thank M. Reuter and S. Theisen for
helpful discussions. C. S. also thanks J. Lauer for
discussions and the
Deutsche Forschungsgemeinschaft for financial
support during part of this work.
\vfill\eject
\vskip1cm
\hskip3cm

{\bf References}

\vskip.5cm
\parindent=17pt

\item{[1]}{Z. Bern and D. A. Kosower,
Phys. Rev. Lett. {\bf 66} (1991) 1669;
Nucl. Phys. {\bf B379} (1992) 451.}
\item{[2]}{Z. Bern and D. C. Dunbar, Nucl. Phys. {\bf B379} (1992) 562.}
\item{[3]}{Z. Bern, L. Dixon and D. A. Kosower, SLAC-PUB-5947;
           Z. Bern, D. C. Dunbar and T. Shimada, Phys. Lett. {\bf B312}
           (1993) 277;
           Z. Bern, UCLA-93-TEP-5 and references therein.}

\item{[4]}{M. J. Strassler, Nucl. Phys. {\bf B385} (1992) 145.}
\item{[5]}{M. J. Strassler, SLAC-PUB-5978.}
\item{[6]}{J. Schwinger, Phys. Rev. 82 (1951) 664.}
\item{[7]}{C. Itzykson and J. Zuber, {\sl Quantum Field Theory},
           McGraw-Hill 1980.}

\item{[8]}{A. O. Barvinsky, G. A. Vilkovisky,
Nucl. Phys. {\bf B282} (1987) 163;
Nucl. Phys. {\bf B333} (1990) 471; Nucl. Phys. {\bf B333} (1990) 512;
G. A. Vilkovisky, CERN-TH 6392/92;
A. O. Barvinsky, Yu. V. Gusev, V. V. Zhytnikov, G. A. Vilkovisky,
MANITOBA-93-0274.}
\item{[9]}{L. Alvarez-Gaum\'e, Commun. Math. Phys. {\bf 90} (1983) 161.}
\item{[10]}{M. G. Schmidt and C. Schubert, in preparation.}
\item{[11]}{M. A. Shifman, Nucl. Phys. {\bf B173} (1980) 12,
and references therein.}
\item{[12]}{L. Brink, P. Di Vecchia, P. Howe,
Nucl. Phys. {\bf B118} (1977) 76;
A. P. Balachandran, P. Salomonson, B. Skagerstam and J. Winnberg,
Phys. Rev. {\bf D15} (1977) 2308;
F. A. Berezin and M. S. Marinov, Ann. Phys. (NY) {\bf 104} (1977) 336;
R. Casalbuoni, Phys. Lett. {\bf B62} (1976) 49;
A. Barducci, R. Casalbuoni and L. Lusanna, Nuovo Cim. {\bf 35A} (1976) 377;
J. W. van Holten, in: {\sl Proc. Sem. Math. Structures in Field Theories}
(1986-87), CWI syllabus vol. 26 (1990) 109.}
\item{[13]} {L. Carson, L. McLerran, Phys. Rev. {\bf D41} (1990) 647.}
\item{[14]} {L. Carson, Phys. Rev. {\bf D42} (1990) 2853.}
\item{[15]} {R. R. Metsaev and A. A. Tseytlin,
             Nucl. Phys. {\bf B298} (1988) 109.}

\bye